\documentclass[conference]{IEEEtran}
\IEEEoverridecommandlockouts

\usepackage{booktabs}
\usepackage{multirow}
\usepackage[frozencache,cachedir=minted-cache]{minted}

\usepackage[many]{tcolorbox}
\usepackage{hyperref}

\newcommand{\area}{
    \begin{minipage}[b]{0.06\columnwidth}
        \centering
        \raisebox{-.05in}{\includegraphics[width=\linewidth]{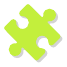}}
    \end{minipage}
}

\newcommand{\perf}{
    \begin{minipage}[b]{0.06\columnwidth}
        \centering
        \raisebox{-.05in}{\includegraphics[width=\linewidth]{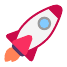}}
    \end{minipage}
}

\newcommand{\power}{
    \begin{minipage}[b]{0.06\columnwidth}
        \centering
        \raisebox{-.05in}{\includegraphics[width=\linewidth]{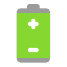}}
    \end{minipage}
}

\newcommand{\programmer}{
    \begin{minipage}[b]{0.06\columnwidth}
        \centering
        \raisebox{-.05in}{\includegraphics[width=\linewidth]{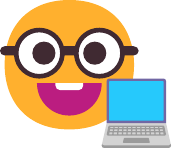}}
    \end{minipage}
}

\newcommand{\reviewer}{
    \begin{minipage}[b]{0.06\columnwidth}
        \centering
        \raisebox{-.05in}{\includegraphics[width=\linewidth]{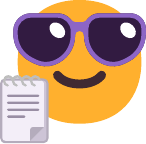}}
    \end{minipage}
}

\newcommand{\analyst}{
    \begin{minipage}[b]{0.06\columnwidth}
        \centering
        \raisebox{-.05in}{\includegraphics[width=\linewidth]{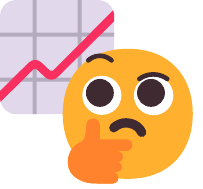}}
    \end{minipage}
}

\newcommand{\observer}{
    \begin{minipage}[b]{0.06\columnwidth}
        \centering
        \raisebox{-.05in}{\includegraphics[width=\linewidth]{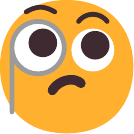}}
    \end{minipage}
}

\newcommand{\checked}{
    \begin{minipage}[b]{0.06\columnwidth}
        \centering
        \raisebox{-.05in}{\includegraphics[width=\linewidth]{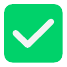}}
    \end{minipage}
}

\tcbuselibrary{minted}
\newtcblisting{colorcode}{
    colframe=lightgray,
    colback=lightgray!10,
    boxrule=0.5pt,
    listing only,
    listing engine=minted,
    minted language=verilog,
    minted options={fontsize=\scriptsize},
    height fixed for=all
}

\makeatletter
\newcommand{\linebreakand}{
    \end{@IEEEauthorhalign}
    \hfill\mbox{}\par
    \mbox{}\hfill\begin{@IEEEauthorhalign}
}
\makeatother

\def\BibTeX{{\rm B\kern-.05em{\sc i\kern-.025em b}\kern-.08em
    T\kern-.1667em\lower.7ex\hbox{E}\kern-.125emX}}
\begin{document}

\title{RTLSquad: Multi-Agent Based \\Interpretable RTL Design}
\author{
    \IEEEauthorblockN{1\textsuperscript{st} Wang Bowei}
    \IEEEauthorblockA{
        \textit{Computer Department} \\
        \textit{National University of Defense Technology}\\
        Changsha, China\\
        wangbowei@nudt.edu.cn
    }
    \and
    \IEEEauthorblockN{2\textsuperscript{nd} Qi Xiong}
    \IEEEauthorblockA{
        \textit{Computer Department} \\
        \textit{National University of Defense Technology}\\
        Changsha, China\\
        xiongqi@nudt.edu.cn
    }
    \linebreakand
    \IEEEauthorblockN{3\textsuperscript{rd} Zeqing Xiang}
    \IEEEauthorblockA{
        \textit{Intelligent Microelectronics Center} \\
        \textit{Qiyuan Lab}\\
        Beijing, China\\
        xiangzeqing20@mails.ucas.ac.cn
    }
    \and
    \IEEEauthorblockN{4\textsuperscript{th} Lei Wang}
    \IEEEauthorblockA{
        \textit{Defense Innovation Institute} \\
        \textit{Academy of Military Science}\\
        Beijing, China\\
        leiwang@nudt.edu.cn
    }
    \and
    \IEEEauthorblockN{5\textsuperscript{th} Renzhi Chen}
    \IEEEauthorblockA{
        \textit{Intelligent Microelectronics Center} \\
        \textit{Qiyuan Lab}\\
        Beijing, China\\
        chenrenzhi@qiyuanlab.com
    }
}

\maketitle

\begin{abstract}
    Optimizing Register-Transfer Level (RTL) code is crucial for improving hardware PPA performance. Large Language Models (LLMs) offer new approaches for automatic RTL code generation and optimization. However, existing methods often lack decision interpretability (sufficient, understandable justification for decisions), making it difficult for hardware engineers to trust the generated results, thus preventing these methods from being integrated into the design process. To address this, we propose RTLSquad, a novel LLM-Based Multi-Agent system for interpretable RTL code generation. RTLSquad divides the design process into exploration, implementation, and verification \& evaluation stages managed by specialized agent squads, generating optimized RTL code through inter-agent collaboration, and providing decision interpretability through the communication process. Experiments show that RTLSquad excels in generating functionally correct RTL code and optimizing PPA performance, while also having the capability to provide decision paths, demonstrating the practical value of our system.

\end{abstract}

\begin{IEEEkeywords}
    IC Design, Large Language Model, Multi-Agent
\end{IEEEkeywords}

\section{Introduction}
\label{sec:introduction}

\begin{figure}[ht]
    \centering
    \includegraphics[width=0.90\linewidth]{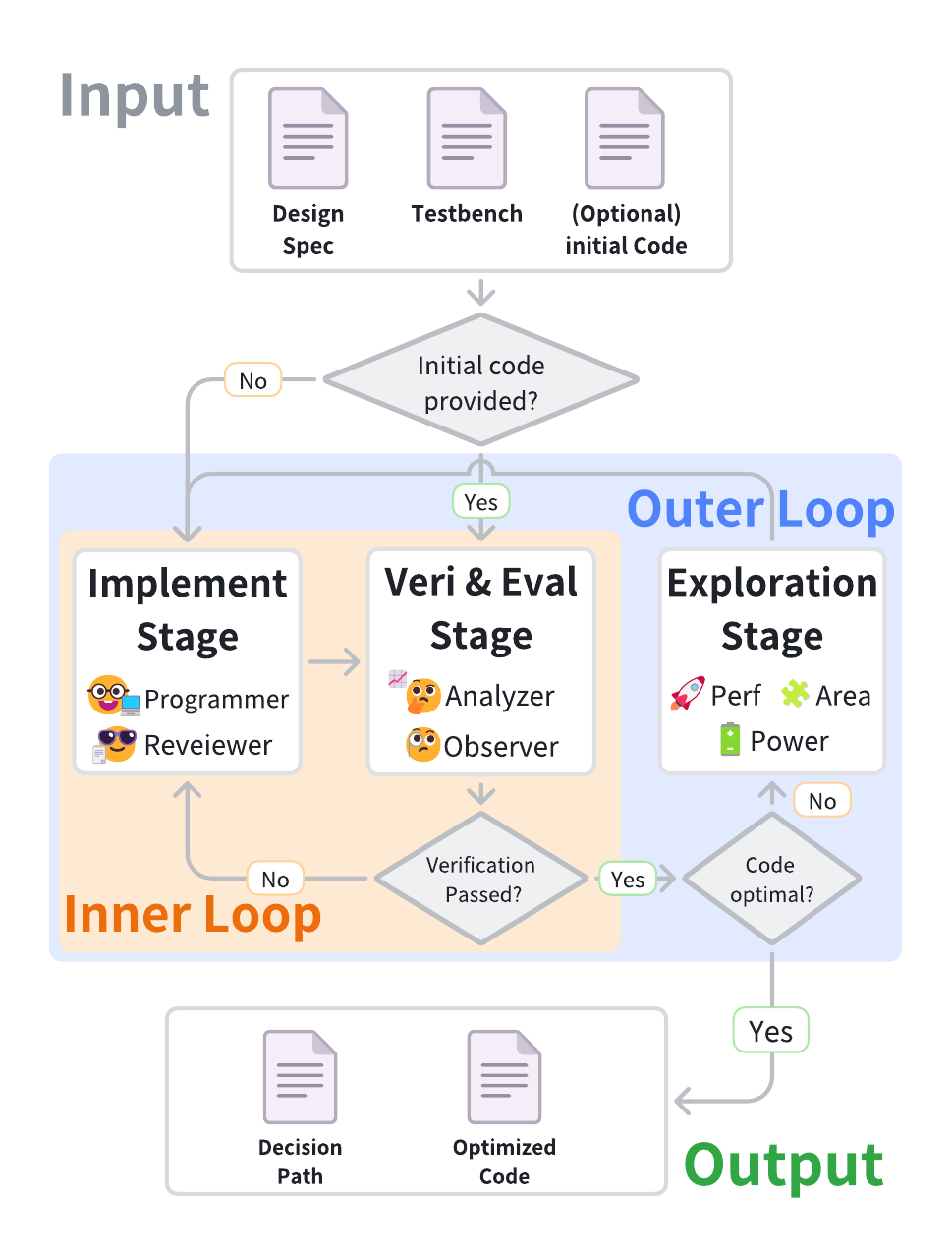}
    \caption{The overview of RTLSquad. The workflow consists of three stages: exploration, implementation, and verification \& evaluation. The iterative process is divided into inner loop and outer loop based on verification results.}
    \label{fig:overview}
\end{figure}

Register-transfer level (RTL) design fundamentally impacts the power, performance, and area (PPA) of hardware, highlighting the importance of optimizing RTL code. Recent research has introduced large language model (LLM) techniques for automated RTL code generation, either by fine-tuning to produce high-quality RTL code~\cite{liuRTLCoderFullyOpenSource2024,thakur2022benchmarkinglargelanguagemodels,thakurVeriGenLargeLanguage2024} or through multi-stage processes to progressively optimize RTL code~\cite{yaoRTLRewriterMethodologiesLarge2024,delorenzoMakeEveryMove2024}.

Multi-stage RTL code generation process requires decision-making ability (what design to select, how to implement the design, etc.). Existing work leverages traditional heuristic algorithms to guide LLMs in making decisions, such as Monte Carlo Tree Search algorithm~\cite{yaoRTLRewriterMethodologiesLarge2024,delorenzoMakeEveryMove2024,LLMCodeGen}. These methods can provide \textit{result interpretability} by providing generated code after each decision, but they are limited in providing \textbf{decision interpretability}~\cite{krausAIExplainingDecisions2020}, the ability to clearly explain the rationale behind each decision. This limitation arises from the inherent randomness and opaque nature of heuristic methods. The lack of decision interpretability leads to: 1) Hardware engineers may struggle to trust the generated results, therefore unlikely to integrate these methods into current design process~\cite{mohammadkhani2023systematicliteraturereviewexplainable}; 2) Engineers being unable to discern the quality of results during the generation process, thus lacking sufficient reason to accept the final result over others; 3) Engineers also being unable to learn from the decision-making process, therefore possible to adopt an erroneous design found to be faulty in the past again.

The LLM-based multi-agent (LLM-MA) system is an innovative method for providing decision interpretability. LLM-MA system consists of multiple LLM agents with different role settings, solve tasks by simulating human collaboration~\cite{chanChatEvalBetterLLMbased2023,li2024improvingmultiagentdebatesparse,smit2024goingmadlookmultiagent}. This setup demonstrates greater problem-solving capabilities compared to a single LLM instance~\cite{qian2024chatdevcommunicativeagentssoftware,liangEncouragingDivergentThinking2023}. A key feature of the system is that agents communicate and negotiate decisions using natural language, making the decision-making process inherently transparent and comprehensible to users, thereby offers significant advantages in providing decision interpretability.

To fully leverage the advantages of the LLM-MA system in decision interpretability for hardware design, we propose \textbf{RTLSquad}, a novel LLM-MA system dedicated to RTL code generation and optimization. RTLSquad operates through a multi-stage process, as shown in Figure~\ref{fig:overview}: \textit{exploration stage}, exploring various design options to optimize PPA; \textit{implementation stage}, generating correct RTL code according to specifications and requirements; \textit{verification \& evaluation stage}, analyzing EDA tool reports and identifying trends in data changes. Each stage is driven by a squad composed of multiple specialized LLM agents, where the communication process within the squad can be regarded as an interpretation of each design decision step, thereby providing decision interpretability for hardware engineers. Through iterative optimization, RTLSquad can progressively explore and implement various designs based on tool feedback, ultimately producing RTL code that is both correct and well-optimized.

Experimental results across various designs demonstrate that RTLSquad's iterative optimization enhances backbone model's ability to generate correct RTL code, with Pass@1 performance improving by 7.2\% and matching or exceeding the PPA of the reference design in most cases (73.3\%). The decision paths are formatted into well-structured documents as outputs, providing comprehensive explanations for each decision in each stage, effectively addressing the disadvantages caused by insufficient decision interpretability. Our contributions are summarized as follows:

\begin{itemize}
    \item We propose RTLSquad, an RTL code generation and optimization framework dedicated to providing reliable and understandable hardware design.
    \item We have developed various interaction mechanisms to enable multiple agents to effectively collaborate on generating RTL code.
    \item We are the first to be able to generate RTL code while providing decision paths, thereby possessing greater practical practical value.
\end{itemize}

\section{Related Work}
\label{sec:related_work}

\subsection{LLM-Based Agent for RTL Design}
\label{sub:related-llm-eda}

LLM for hardware design has gradually gained attention from researchers. Related studies covering various design processes~\cite{kahng2024dgreplacedataflowdrivengpuacceleratedanalytical,hosny2019drillsdeepreinforcementlearning,8367260,TIAN2021174}. Our work primarily focuses on automated RTL design. Relevant studies indicate that LLMs capable of generate Verilog code but with limitations~\cite{liuVerilogEvalEvaluatingLarge2023}. The agent paradigm can enhance LLM's design capabilities, such as using the ReAct~\cite{yao2023reactsynergizingreasoningacting} paradigm to drive LLMs to fix Verilog syntax errors~\cite{tsaiRTLFixerAutomaticallyFixing2024}. Further, we focus on the LLM-MA (LLM-Based Multi Agent) method~\cite{guoLargeLanguageModel2024}, which completes tasks through collaboration among multiple agents. Existing work primarily utilizes LLM-MA for HLS design~\cite{sheikholeslamSynthAIMultiAgent2024} or layout design~\cite{liuLayoutCopilotLLMpoweredMultiagent2024}, lacking attention to RTL design. Our work employs multiple agents to form multiple specialized squads of agents, used to generate and optimize RTL code.

\subsection{Decision Interpretability with LLM-MA}

AI interpretability enables users to understand why AI systems produce specific results, thereby enhancing trust~\cite{SAEED2023110273}. The application of LLM-MA systems in decision interpretation has been achieved in several critical fields such as medical diagnosis~\cite{hong2024argmedagentsexplainableclinicaldecision,pandeyAdvancingHealthcareAutomation2024}, fact-checking~\cite{kim2024llmsproducefaithfulexplanations}, etc. This demonstrates the feasibility of using such methods in hardware design, a field with equally high demands for reliability. Compared to decision paths generated by methods like CoT~\cite{wei2023chainofthoughtpromptingelicitsreasoning}, LLM-MA systems provide superior decision interpretability, including robust justification for decisions~\cite{duImprovingFactualityReasoning2023} and better diversity~\cite{liangEncouragingDivergentThinking2023}. Therefore, our work employs a squad of agents as decision-makers, aiming to provide complete decision interpretations for implementation, exploration, and analysis.

\section{Methodology}
\label{sec:methodology}

In this section, we will comprehensively introduce the design and implementation of RTLSquad. First, we will show the workflow of RTLSquad (Section~\ref{sub:method_overview}). Then, we will introduce the exploration (Section~\ref{sub:method_exploration}), implementation (Section~\ref{sub:method_implementation}), and verification \& evaluation stages (Section~\ref{sub:method_verification}) of RTLSquad, describing the roles and processes of each stage.

\subsection{Overview}
\label{sub:method_overview}

Figure~\ref{fig:overview} illustrates the workflow of RTLSquad. Consistent with the standard RTL design process, RTLSquad accepts design specifications, testbench files, and optional initial code as inputs, and outputs optimized code and comprehensive decision path documentation. Depending on whether the functional verification is passed, the iterative process of RTLSquad is divided into the \textbf{verification-fix loop (inner loop)} and the \textbf{exploration-implementation loop (outer loop)}. If the current version of the RTL code fails the functional verification, it enters the inner loop, switching between the verification and implementation stages to obtain error reports and correct errors in the code; if verification is successful, RTLSquad enters the outer loop, switching between the exploration, implementation, and verification stages to explore and implement new design methods.

\subsection{Implementation Stage}
\label{sub:method_implementation}

\begin{figure}[ht]
    \centering
    \includegraphics[width=0.93\linewidth]{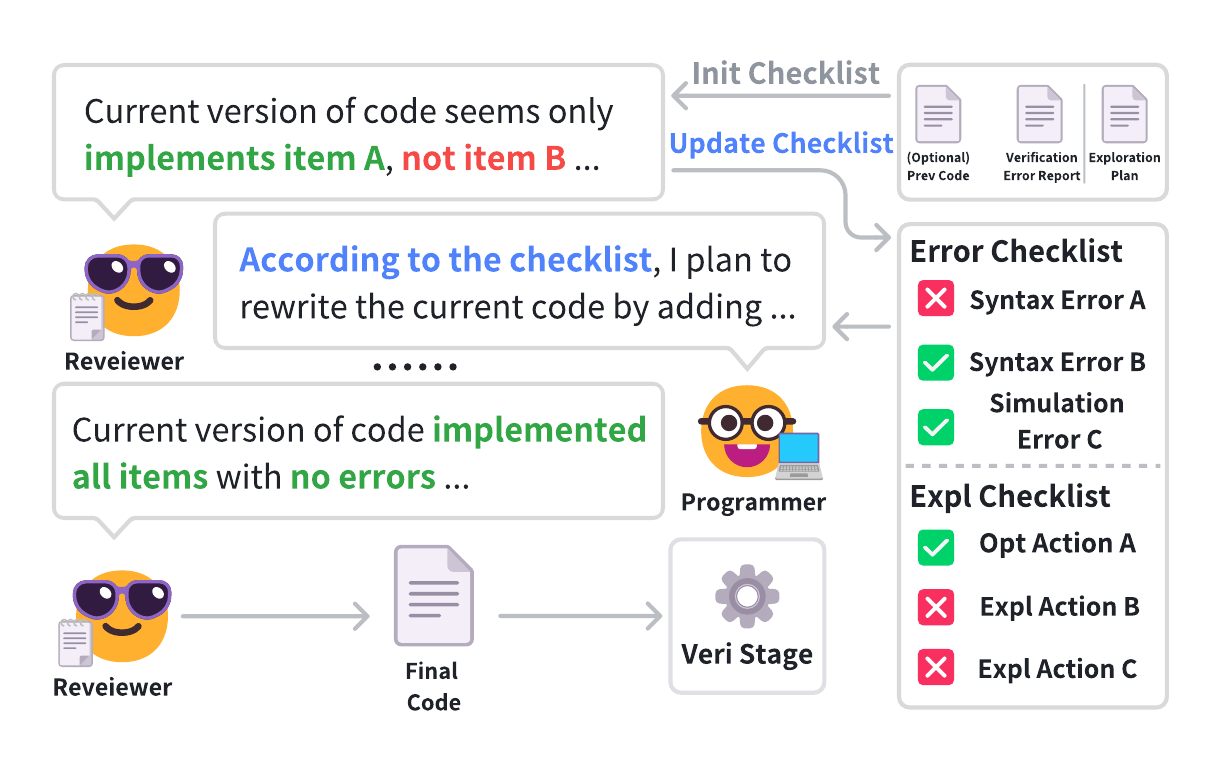}
    \caption{The implementation stage of RTLSquad.}
    \label{fig:impl}
\end{figure}

As shown in Figure~\ref{fig:impl}, the implementation stage is used to implement the exploration plan (see Section~\ref{sub:method_exploration}) or to fix errors discovered during the verification stage. The implementation squad $\mathcal{S}_{Impl}$ consists of a \textbf{programmer} and a \textbf{reviewer}. A checklist is maintained during the operation, with each item representing a task to be implemented or fixed. The programmer first summarizes the incomplete tasks in the checklist and the reviewer's feedback, then plans the code generation steps, and finally generates a new version of the code. The reviewer is responsible for reviewing the code, providing feedback on potential errors to the programmer, and overseeing whether the code completes the tasks in the checklist. The implementation stage will only conclude and proceed to the verification stage when the reviewer marks all tasks as completed. The multi-round feedback and correction process based on review provides better RTL code generation capabilities (see Section~\ref{sub:expr_functional_correctness}).

\subsection{Verification \& Evaluation Stage}
\label{sub:method_verification}

\begin{figure}
    \centering
    \includegraphics[width=0.93\linewidth]{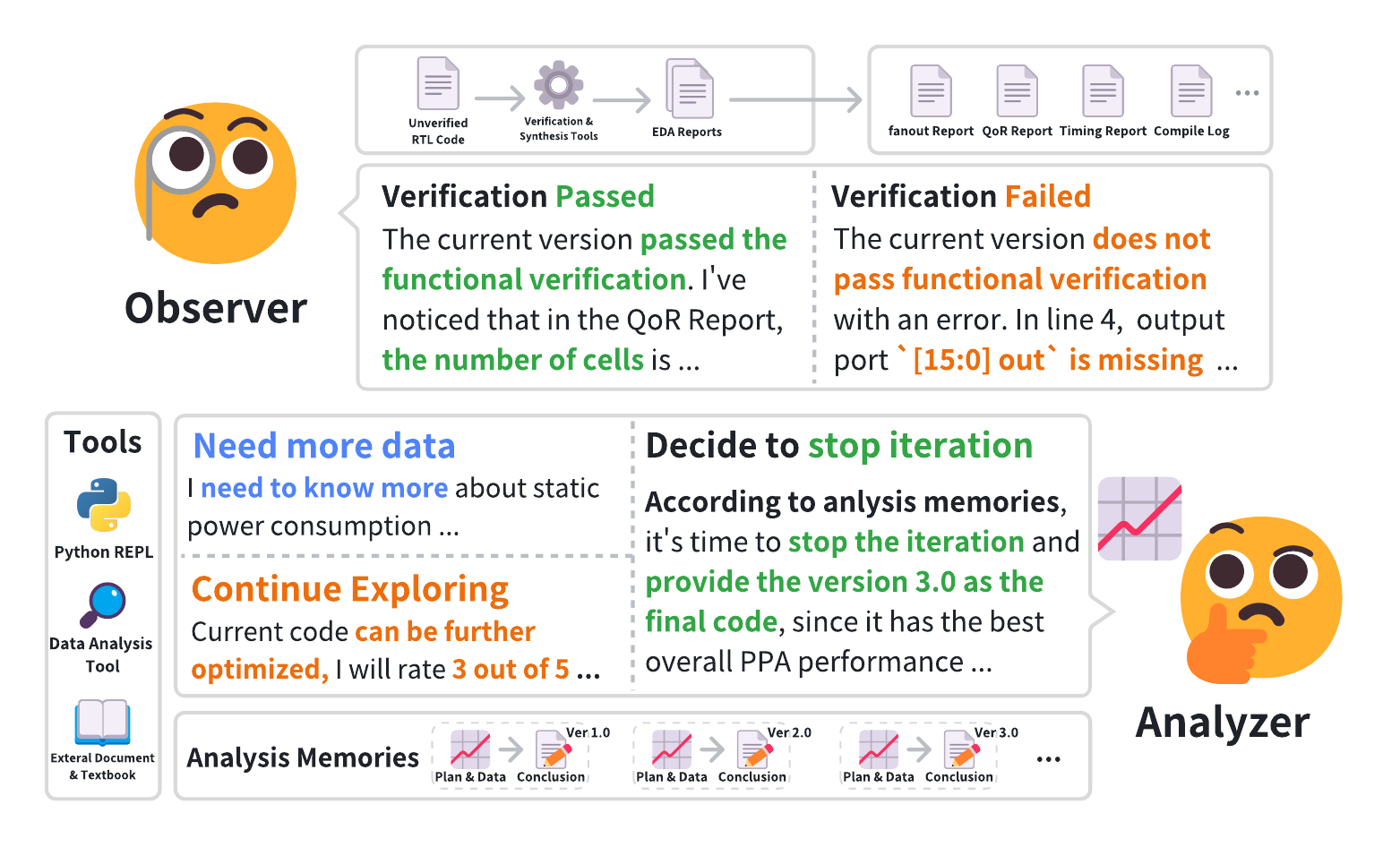}
    \caption{The Verification \& Evaluation stage of RTLSquad.}
    \label{fig:veri}
\end{figure}

As shown in Figure~\ref{fig:veri}, the verification \& evaluation stage is used to perform functional simulation and logic synthesis on the RTL code provided by the implementation stage, observe and analyze the reports, and evaluate the design. The squad $\mathcal{S}_{Veri}$ in the verification stage consists of an \textbf{observer} and an \textbf{analyst}. The observer is responsible for extracting important information such as compilation errors, verification failures, and abnormal performance indicators from the reports and conveying it to the analyst. The analyst analyzes the statistical characteristics of the indicators based on the current and past analysis records of the current code.

The analyst will give a rating score $r^{(k)} \in [1, 5]$ on the exploration plan, representing the effectiveness of the plan in optimizing PPA performance, with a higher score indicating greater effectiveness. The rating results will be used to adjust the exploration points $p^{(k+1)}$ for the next round, guiding the agent to dynamically adjust the decision-making direction based on past performance, calculated as:

\begin{align}
     & \hat{p}^{(k)} = \left(\alpha - \beta r\right) \cdot p^{(k)} \label{1} \\
     & p^{(k + 1)} = \eta \hat{p}^{(k)} + (1 - \eta) p^{(k)}
\end{align}

where $\alpha, \beta$ are the mapping factors from scores to scaling coefficients, typically set as $\alpha = 2.375, \beta = 0.375$, $\eta$ is the smoothing factor, typically set as $\eta = 0.4$. The above calculation results in an inverse relationship between the scores and the number of exploration points for the next round, thereby encouraging progressive optimization or break away from the current design.

The analyst will determine if the current code has reached its optimal state. If further improvements are possible, they will transition to the exploration stage to attempt other designs. If the performance metrics of the code show a flat or declining trend, new exploration actions may not bring performance improvements. The analyst will choose to terminate the iteration, selecting the version with the overall best performance from all historical code versions as the output. Subsequently, the user can decide whether to accept this version of the RTL code or continue optimizing.

\subsection{Exploration Stage}
\label{sub:method_exploration}

\begin{figure}
    \centering
    \includegraphics[width=0.93\linewidth]{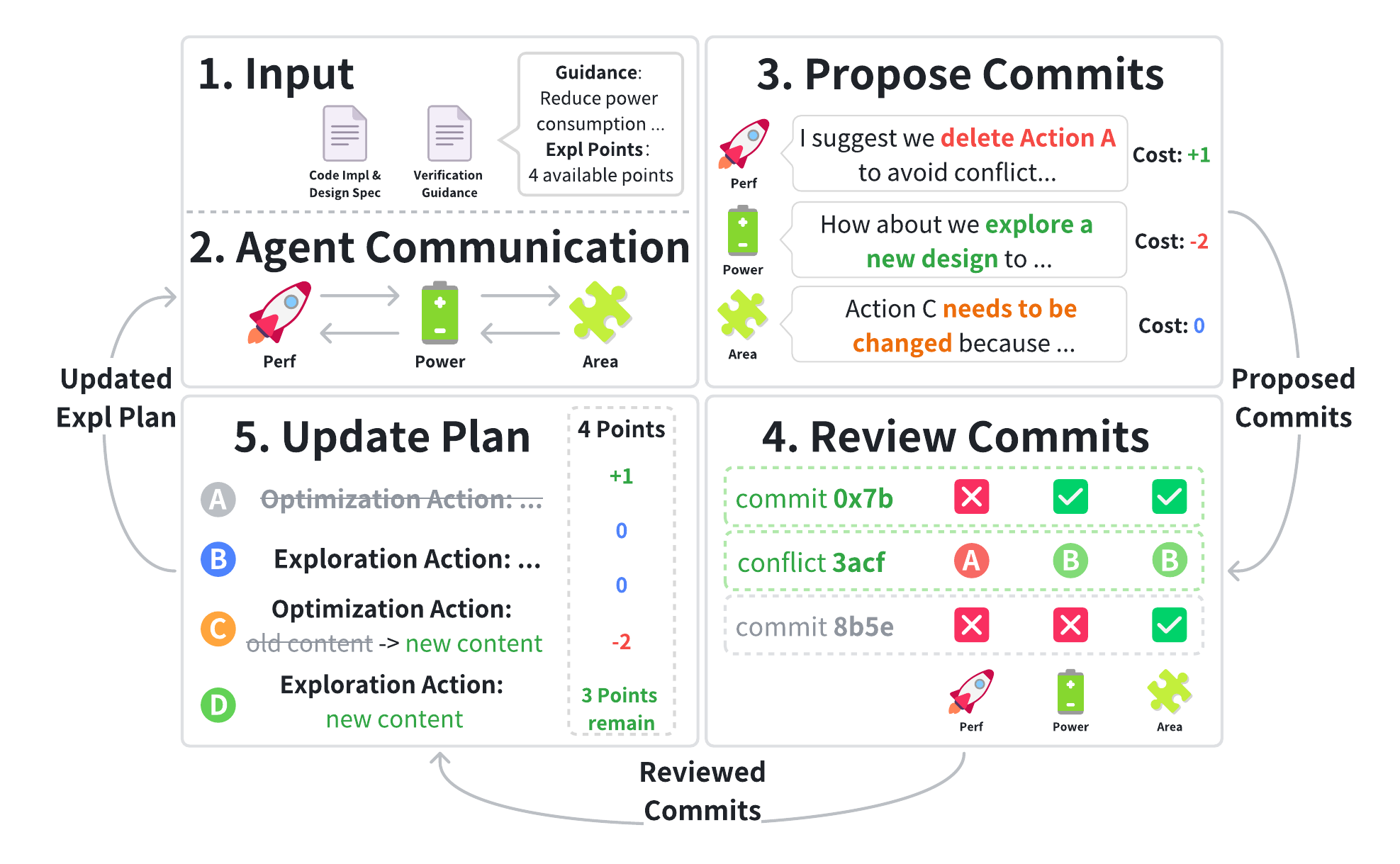}
    \caption{The exploration stage of RTLSquad.}
    \label{fig:expl}
\end{figure}

As shown in Figure~\ref{fig:expl}, the exploration stage is used to explore different designs, forming a set of code design decisions that can optimize current code in terms of PPA. The exploration stage includes a squad $\mathcal{S}_{expl}$ composed of \textbf{experts in optimizing power, performance, and area}, where three agents will engage in a debate. The agents communicate through a shared message pool~\cite{chanChatEvalBetterLLMbased2023}. The three agents take turns speaking, and the order of speaking in each round will be randomly arranged to ensure fairness.

During the debate, an exploration plan consisting of exploration actions (code design decisions to implement in the implementation stage) is maintained. First, agents will \textit{communicate} with each other, discussing current exploration plan. Then, agents will propose \textit{commits}, including adding new actions or modifying/deleting existing actions. Finally, agents will \textit{review} the commits, judging their acceptability based on their own stance. Commits that receive a majority vote will be applied to the exploration plan. When all agents have not proposed new commits and the exploration plan is acceptable (see below), the exploration stage will end, transitioning to the implementation stage.

Exploration actions are restricted to exploration (\texttt{expl}) and optimization (\texttt{opt}), representing radical or incremental modifications to the existing design. The number and type of actions are limited by the \title{exploration points} $p^{(k)}$, where \texttt{expl} actions consume 2 points and \texttt{opt} actions consume 1 point. Exploration points can heuristically guide the agent to dynamically adjust the aggressiveness of exploration (proposing more exploration or optimization actions) based on past performance, while not imposing excessive constraints to ensure flexibility. The available points for each round are provided by the verification stage (detailed in Section~\ref{sub:method_verification}).

\section{Case Study: Decision Interpretability}
\label{sec:case-study}

\begin{table}[ht!]
  \label{tab:task1}

    \centering
    \begin{tabular}{p{0.9\linewidth}}
        \toprule
        \textbf{Example A: Implementation Interpretability} \\
        \midrule
      \begin{center}
        \textbf{User requests to design an 8-bit data accumulator.} \\
        \programmer is programmer, \reviewer is reviewer. 
      \end{center}

        \programmer: Exploration squad suggested refactoring the accumulation logic to reduce the critical path delay, to address this, I will follow a structured implement steps: 1. Define a accumulator register ... 2. Implement Count logic ... \\
        \programmer: The code implementation are as follows: \\

\begin{colorcode}
module accu(/* ports */);
  // Single accumulator register
  reg [9:0] accumulator; 

  always @(posedge clk or negedge rst_n) begin
  // ... count logic
  end

  always @(posedge clk or negedge rst_n) begin
  // ... accumulation logic
  end

  always @(posedge clk or negedge rst_n) begin
  // ... Output logic
  end
endmodule
\end{colorcode}
        \programmer: @reviewer, I have rewritten the code ...   \\
        \reviewer: @programmer, please address the following issues in the code: The output logic should be updated to ensure that \texttt{valid\_out} is set to 1 only when the accumulation of four valid inputs is complete. Currently,  \texttt{valid\_out} is set to 1 whenever \texttt{valid\_in} is high and \texttt{count} is 3, which does not fully align with the design specification... \\
        \programmer: To address the feedback and ensure the code meets all requirements, I will follow these steps: ...                                                                 \\
\begin{colorcode}
module accu(/* ports */);
  // Single accumulator register
  reg [9:0] accumulator; 
  // Combined count and accumulation logic
    always @(posedge clk or negedge rst_n) begin
      if (!rst_n) begin
      // ... init
      end else if (valid_in) begin
      // ... set accumulator
      end else begin
        valid_out <= 0;
      end
    end
endmodule
\end{colorcode}
        \programmer: @reviewer, I have addressed the feedback ... The count and accumulator are now updated within a single \texttt{always} block ... If there are any further suggestions or concerns, please let me know. \\
        \bottomrule
    \end{tabular}

\end{table}

\begin{table}[ht]
    \label{tab:task2}

    \centering
    \begin{tabular}{p{0.9\linewidth}}
        \toprule
        \textbf{Example B: Verification Interpretability} \\
        \midrule
        \begin{center}
            \textbf{User requests to design a Syncronizer.} \\
            \observer is observer, \analyst is analyst.
        \end{center}
        \observer : The QoR report indicates that the design has no timing violations, with a critical path length of 0.38 ns. The design consists of 17 leaf cells, with 6 combinational and 11 sequential cells ... \\
        \analyst : The current design shows good performance with no timing violations and a relatively small area footprint. However, to fully evaluate the effectiveness of the exploration plan, additional data on power consumption is needed. \\
        \analyst : \textbf{Please provide the power consumption data} for the "synchronizer" design, including static and dynamic power metrics. Specifically, we need the total power consumption, breakdown of static and dynamic power, and any power-related violations or warnings.
        \begin{tcolorbox}[colframe=lightgray,colback=lightgray!10,boxrule=0.5pt]
            \centering
            \observer Reading the \texttt{power\_report} file to extract the power consumption data for the "synchronizer" design.
        \end{tcolorbox}
        \observer : The power report for the "synchronizer" design indicates a total power consumption of 4.6652e-03 mW, with 28.0450 nW static power and 4.6372 uW dynamic power. There are no power-related violations or warnings in the design. \\
        \analyst : Based on memories, the \textbf{previous version had low timing performance}. The \textbf{current version uses the asynchronous FIFO synchronization recommended by the exploration team} to reduce the latency of synchronization across clock domains, which has performed well, and thus I give the current design solution a rating of 4/5. In the next optimization, I would like the design team to \textbf{explore new design that reduces the power consumption} ...
        \\\bottomrule
    \end{tabular}

\end{table}

In this section, we analyze the results generated by RTLSquad under three different typical designs to demonstrate decision interpretability at each stage and explain how these interpretabilities can address the issues inherent in existing methods. The content displayed is excerpted from the output decision path documentation and has been reformatted for readability. Due to space limitations, the examples provided here do not cover all design scenarios, but are sufficient to illustrate the forms of interpretability provided by RTLSquad.

\subsection{Implementation Interpretability}
\label{sub:case-study-impl}

In Example~\ref{tab:task1}, the programmer first listed the implementation steps based on the exploration plan, and then implemented them step by step. During the implementation process, the programmer initially provided a syntax-correct implementation that met the exploration plan but did not conform to the design specifications. The reviewer, after reviewing, found that the current code met the requirements of the exploration plan but contained functional errors, and thus made modification suggestions to the programmer. Under the reviewer's suggestions, the programmer made targeted modifications and refactoring to the code, ultimately providing the correct implementation.

\textbf{Conclusion}. The decision-making process during the implementation stage includes steps for RTL code generation, issues within the code, and the code refactoring actions required to address these issues. This information provides transitional details before and after code modifications, decomposing the otherwise difficult-to-understand code changes into specific steps and objectives, thereby reducing the difficulty of understanding. Thus, through the decision interpretability of the implementation stage, hardware engineers can more easily review the RTL code generated by RTLSquad, resulting in higher confidence in the generated results and their reliable integration into the existing design process.

\subsection{Analysis Interpretability}
\label{sub:case-study-analysis}

In Example~\ref{tab:task2}, after performing logic synthesis, the observer first read the QoR report, extracted key performance indicators such as clock frequency, area, and power consumption, and then informed the analyst. The analyst summarized the data and gave clear instructions to the observer, requesting more data on power consumption. After the observer provided the data, the analyst referred to the previous round's analysis records and the current version's code implementation, analyzed the data trends, and concluded that the new design adopted in the current version effectively reduced latency, thus deemed effective; however, the code could still be further optimized. Based on the conclusion, the analyst outlined the focus areas for the next round of exploration, requiring optimization aimed at reducing power consumption.

\begin{table}[ht]
    \label{tab:task3}

    \centering
    \begin{tabular}{p{0.9\linewidth}}
        \toprule
        \textbf{Example C: Exploration Interpretability} \\
        \midrule
        \begin{center}
            \textbf{User requests to optimize a pipelined 64-bit ripple carry adder.} \\
            \area is area, \power is power, \perf is performance.
        \end{center}

        \textit{Previous steps omitted for brevity.}  \\

        \power : I suggest we \textbf{implement clock gating in the pipeline stages} to reduce dynamic power consumption. Specifically, gate the clock for stages that do not need to be updated in every cycle. Clock gating is a well-known technique to reduce dynamic power consumption by preventing unnecessary clock toggling. This will be particularly effective in reducing power in idle or less frequently used pipeline stages. This action will consume 1 exploration points. \\

        \area : @power, I agree with your proposal to implement clock gating in the pipeline stages to reduce dynamic power consumption. This is a solid optimization action that aligns well with our goal of reducing power without compromising performance.

        \area : Also, I suggest we try to \textbf{reduce the number of registers} in the pipeline stages by \textbf{reusing intermediate registers} and simplifying the pipeline structure. This action will help in reducing the area footprint of the design without significantly impacting performance. By reusing registers and simplifying the pipeline, we can achieve a more compact design. This action will consume 2 exploration points.

        \begin{tcolorbox}[colframe=lightgray,colback=lightgray!10,boxrule=0.5pt,title={Current Commits}]
            \textbf{Commit 1} (\texttt{79b4}): Optimization Action: Implement clock gating in the pipeline stages to reduce dynamic power consumption. \\
            \textbf{Commit 2} (\texttt{982d}): Exploration Action: Reduce the number of registers in the pipeline stages by reusing intermediate registers and simplifying the pipeline structure.
        \end{tcolorbox}

        \textit{Other reviews omitted for brevity.}  \\

        \perf : For Commit \texttt{79b4}, on one hand, clock gating is a well-known technique to reduce dynamic power consumption by preventing unnecessary clock toggling. This will be particularly \textbf{effective in reducing power in idle or less frequently used pipeline stages}; on the other hand, Implementing clock gating might introduce additional complexity in the design and could potentially affect timing if not managed properly. I believe we can handle this complexity if we work together with the verification and implementation teams, so I suggest we \textbf{accept this commit}. \\
        \perf: For Commit \texttt{982d}, however, refactoring the pipeline stages to use a single register for intermediate results will help in reducing the area footprint of the design without significantly impacting performance. But this action might introduce timing issues if not carefully managed, as it reduces the number of registers and \textbf{could lead to data hazards}. To avoid potential timing issues, I suggest we \textbf{reject this commit} and explore other optimization actions.
        \begin{tcolorbox}[colframe=lightgray,colback=lightgray!10,boxrule=0.5pt]
            \centering
            \checked Commit \texttt{79b4} Accepted, add a new action into Exploration plan.
        \end{tcolorbox}
        \\\bottomrule
    \end{tabular}

\end{table}

\textbf{Conclusion}. The decision-making process in the verification \& evaluation stage has presented the evaluation process for the current version of the RTL code. The evaluation process comprehensively considers the PPA data provided in the EDA tool reports, the design decisions implemented in the current code, and the historical analysis results contained in the Analysis Memories, which serve as the evidence for the evaluation process, ultimately leading to a certain type of evaluation result. Through the decision interpretability of the verification \& evaluation stage, engineers can understand the basis and logic of the evaluation process, thus having a clear perception of the quality of a certain version of the code, providing sufficient reasons for choosing between code versions.

\begin{table*}[ht]
    \centering
    \footnotesize
    \caption{Comparisons of the PPA performance with reference implementation.}
    \label{tab:rq2}
    \begin{tabular}{c | c c c | c c c | c c c}
        \toprule
        \multirow{2}{*}{Design} & \multicolumn{3}{c|}{Reference} & \multicolumn{3}{c|}{Deepseek} & \multicolumn{3}{c}{RTLSquad-Deepseek}                                                                                                     \\

                                & Power ($\mu W$)                & Perf. ($ns$)                  & Area ($\mu m^2)$                      & Power & Perf.       & Area            & Power           & Perf.                & Area             \\

        \midrule
        adder\_8bit             & 25.86                          & 3.88                          & 40.22                                 & -     & -           & -               & \textbf{24.61}  & \textbf{2.07}        & 40.22            \\
        adder\_16bit            & 52.32                          & 6.92                          & 80.44                                 & 23.57 & 1.49        & 31.36           & \textbf{12.28}  & \textbf{1.26}        & \textbf{20.11}   \\
        adder\_pipe\_64bit      & 187.54                         & \textbf{3.20 (6.60)}          & 2201.04                               & -     & -           & -               & \textbf{102.84} & 8.63 (1.07)          & \textbf{1738.28} \\
        multi\_16bit            & 44.32                          & 8.47 (1.33)                   & 913.55                                & 44.16 & 8.47 (1.33) & \textbf{904.21} & \textbf{42.51}  & \textbf{5.99 (3.81)} & 1133.80          \\
        multi\_booth\_8bit      & 25.74                          & 3.25 (6.55)                   & 412.01                                & 25.74 & 3.25 (6.55) & 412.01          & \textbf{25.67}  & 3.25 (6.55)          & 412.01           \\
        multi\_pipe\_4bit       & 11.51                          & 1.22 (8.56)                   & 163.03                                & 11.51 & 1.22 (8.56) & 163.03          & \textbf{11.43}  & 1.22 (8.56)          & \textbf{161.34}  \\
        pe                      & \textbf{194.13}                & 5.85 (3.95)                   & 2554.4                                & -     & -           & -               & 205.16          & 5.85 (3.95)          & \textbf{2335.38} \\
        pulse\_detect           & \textbf{1.06}                  & 0.94 (8.76)                   & \textbf{15.32}                        & -     & -           & -               & 2.08            & \textbf{0.81 (8.98)} & 29.45            \\
        right\_shifter          & 3.79                           & 0.53 (9.17)                   & 30.64                                 & 3.79  & 0.53 (9.17) & 30.64           & -               & -                    & -                \\
        accu                    & 12.38                          & 2.20 (7.61)                   & \textbf{134.78}                       & 13.67 & 2.03 (7.78) & 192.24          & \textbf{11.76}  & \textbf{2.01 (7.74)} & 192.00           \\
        width\_8to16            & \textbf{14.26}                 & 1.28 (8.52)                   & 176.68                                & 14.53 & 1.83 (7.94) & 198.46          & -               & -                    & -                \\
        \bottomrule
    \end{tabular}
\end{table*}

\subsection{Exporation Interpretability}
\label{sub:case-study-expl}

In Example~\ref{tab:task3}, power experts suggested inserting an optimization design decision (clock gating) to reduce power consumption; area experts proposed to insert an exploration action, restructuring the pipeline stages to reduce the number of registers. Performance experts evaluated both commits, analyzing their advantages and disadvantages, and decided whether to accept or reject based on the current design and potential risks. Finally, the team accepted the first commit, added it to the exploration plan; rejected the second commit, and provided sufficient reasons.

\textbf{Conclusion}. The decision-making process during the exploration stage involves the evaluation of various design decisions. Specifically, when performing the proposal operation, the agent will provide reasons for the modification decisions, and the review stage will critique each modification decision, thus each modification decision will be both defended and challenged. Through the decision interpretability provided by the exploration stage, engineers can understand the advantages and disadvantages of each code modification decision, learning from experience in a comparative manner, thereby avoiding the adoption of those modification decisions that failed to pass the challenge in the design, and preferentially adopting those decisions proven effective after defense.

\section{Experiments}
\label{sec:experiments}

\subsection{Setup}

\textbf{Benchmarks and Baselines}. We used the RTLLM V2.0~\cite{liuOpenLLMRTLOpenDataset} dataset for evaluation, which contains 50 designs with varying scenarios and complexity levels. We optimized the testbench files in the dataset, adding more debugging outputs, enabling the implementation stage to utilize verification run logs for code correction. The optimized dataset will be made publicly available. Consistent with RTLLM, we employed the Self-Planning (SP)~\cite{jiangSelfplanningCodeGeneration2023} method as the baseline approach, which requires the LLM to plan before writing RTL code, then generate code based on the planning results.

\textbf{Experimental Setup}. We utilized Deepseek-V2.5~\cite{deepseekv2} and LLaMA 3.1-70B~\cite{dubey2024llama} as the backbone model for driving agents, with temperature set to 0.8. We used a commercial EDA tool to compile the RTL code and perform functional verification, with testbench files provided by the dataset. For designs that passed verification, we used a synthesis tool equipped with a 40nm process library to perform logic synthesis, obtaining evaluation metrics from output logs.

\textbf{Evaluation Metrics}. We mainly focus on the following categories of metrics: 1. Whether the generated RTL code can be correctly compiled and pass functional verification; 2. Whether the PPA are better than the baseline implementation. For functional verification, we use the Pass@k~\cite{chen2021evaluatinglargelanguagemodels} metric commonly used in code evaluation. Specifically, we focus on the Pass@1 metric, considering the higher cost of hardware code errors in practical scenarios. For PPA, we utilize: Power: dynamic power consumption, in unit $\mu W$; Area: total cell area, in unit $\mu m^2$; Performance: critical path length and slack, in unit $ns$.

\subsection{Functional Correctness}
\label{sub:expr_functional_correctness}

\begin{table}[htbp!]
    \caption{The Pass@1 scores for the Deepseek and LLaMA models using Self-Planning and RTLSquad.}
    \label{tab:rq1}
    \centering
    \begin{tabular}{ccccc}
        \toprule
        \textbf{Method} & \textbf{LLaMA-SP}       & \textbf{Deepseek-SP}       \\
        \midrule
        Pass@1          & 47.6                    & 60.8                       \\
        \toprule
        \textbf{Method} & \textbf{LLaMA-RTLSquad} & \textbf{Deepseek-RTLSquad} \\
        \midrule
        Pass@1          & 58.0                    & 72.0                       \\

        \bottomrule
    \end{tabular}

\end{table}

We conducted experiments across multiple hardware designs to evaluate RTLSquad's capability in generating functionally correct RTL code. We employed rigorous evaluation criteria where any generated code that failed to compile or pass functional verification was considered a failure. Table~\ref{tab:rq1} shows the average Pass@1 results of the SP method and RTLSquad. The results indicate that RTLSquad: 1) can correctly implement designs that the baseline method can implement; 2) through verification feedback, has the ability to accomplish designs that the baseline method cannot implement. Therefore, RTLSquad demonstrates better RTL code generation capability, thus serving as a foundation for the effective operation of design exploration.

\subsection{PPA Metrics}
\label{sub:expr_ppa_performance}

We utilized the verified RTL code provided by the RTLLM dataset and the RTL code generated by the base model using the SP method as baselines, comparing them with the code optimized by RTLSquad in terms of PPA metrics. RTLSquad took the reference implementations from the dataset as input, optimized them to evaluate the framework's capability in optimizing PPA metrics. We executed the generated RTL code using commercial EDA tools for logic synthesis and extracted evaluation metric data from the synthesis reports. The experimental results are shown in Table \ref{tab:rq2}, indicating that in most cases, the RTL code optimized by RTLSquad achieved better overall PPA metrics.

\section{Conclusion}
\label{sec:conclusion}

In this paper, we propose a LLM-MA system for RTL code generation and optimization. We set up exploration, implementation, and verification stages, and assign different squads of agents to perform the functions. This framework not only automates the exploration of designs with superior PPA but also provides the reasoning path for design iteration through communication among agents, thereby enhancing decision interpretability on the basis of result interpretability, making the results more acceptable to designers. This work emphasizes the importance of RTL code generation interpretability, a frequently overlooked issue, and can serve as a prototype for the application of LLM-MA methods in this context, providing a new direction for future research.

\clearpage
\newpage
\bibliographystyle{IEEEtran}
\bibliography{IEEEabrv,main}
\end{document}